\documentclass[11pt]{article}
\usepackage{authblk}
\usepackage{graphicx}
\usepackage{hyperref}
\pdfoutput=1

\title{GeneZip: A software package for storage-efficient processing of genotype data}
\author[1]{Palmer, Cameron\thanks{Correspondence to: cdp2130@columbia.edu}}
\author[1]{Pe'er, Itsik}
\affil[1]{Center for Computational Biology and Bioinformatics, Columbia University Medical Center, New York, NY, USA}

\begin{document}

\maketitle

\begin{abstract}
\section{Summary:}
Genome-wide association studies directly assay $10^6$ single nucleotide polymorphisms (SNPs) across a study cohort.  Probabilistic estimation of additional sites by genotype imputation can increase this set of variants by 10- to 40-fold.  Even with modest sample sizes ($10^3-10^4$), these resulting ``imputed'' datasets, containing $10^{10}-10^{11}$ double-precision values, are incompatible with simultaneous lossless storage in RAM using standard methods.  Existing solutions for this problem require compromises in either genotype accuracy or complexity of permissible statistical methods.  Here, we present a C/C++ library that dynamically compresses probabilistic genotype data as they are loaded into memory.  This method uses a customization of the DEFLATE (gzip) algorithm, and maintains constant-time access to any SNP.  Average compression ratios of $>9-$fold are observed in test data.

\section{Availability and Implementation:}
GeneZip is implemented in C/C++ and relies solely on standard C/C++ libraries; the zlib library is optional for reading compressed genotypes from file.  The beta release for GeneZip is available at \url{http://www.columbia.edu/~cdp2130/genezip}.

\end{abstract}

\section{Introduction}

Over the last decade, SNP arrays have been a primary tool for the study of common human variation.  Array technology
continues to be the method of choice for affordable genome-wide analysis of large cohorts, even in the era of high
throughput (``next generation'') sequencing data, which augment and empower it, rather than replace it.
The standard setting for a Genome-Wide Association Study (GWAS) is array-typing of very large ($10^4-10^6$ individuals) cohorts
across $\sim10^6$ SNPs, augmented by many more variants whose genotypes are imputed based on a reference set of disjoint samples,
such as the 1000 Genomes Project (\cite{Abecasis12}), with denser genotype information, usually from sequencing data.

Genotype imputation is a procedure of statistical inference yielding probabilistic genotypes for study samples at untyped SNPs.
Whereas standard genotyping of a biallelic SNP yields a count of a reference allele at the site, genotype imputation
generates a probability of each of the three possible genotypes.  These probabilities are not immediately compatible with
most common statistical methods used in association studies (see, for example, \cite{Purcell07},
\cite{Kang10}, \cite{Price09}).  Thus, in common practice, probabilistic genotypes are converted
to estimated counts of alleles (\cite{Purcell07} [PLINK v1.07]).

This conversion process is not optimal: discretized or expected allele counts obscure uncertainty inherent in probabilistic
genotype data.  For example, an estimate with equal probability of each genotype $(\frac{1}{3},\frac{1}{3},\frac{1}{3})$ generates
an expected allele count of 1, equivalent to a high-confidence heterozygote genotype; in downstream statistical applications,
such as regression, there is no longer any way for the statistical algorithm to discern between these cases, introducing
noise into the test statistic.  An example of a more valid approach is that taken by the subroutine ``score'' of the
analysis program SNPTEST (\cite{Marchini10}), in which a custom score test is applied that considers each of the genotypes
for each individual in turn, weighted by the probability of the genotype estimated in imputation.  The extensibility
of this approach, however, is limited by the compromise SNPTEST makes to be able to store genotype data in RAM: it considers
only one SNP at a time, discarding the SNP before loading any others.

Issues that interfere with direct analysis of posterior probabilities include:
\begin{itemize}
\item Standard statistical models require substantial adaptation to properly incorporate probabilistic genotypes; and
\item Storage of probabilities in RAM is more complicated for large imputed datasets.
\end{itemize}

This work addresses the second issue.  For a GWAS of $10^4$ individuals imputed to the full 1000 Genomes panel (of 
approximately $4\cdot 10^7$ SNPs and indels), one must store $4*10^{11}$ genotypes.  Simultaneously storing two probabilities per SNP for each sample at double precision
would require 6.4T RAM, well beyond the 4-8G RAM present in many desktop computers.  The most common workaround for this problem is to simply process one SNP at a time (\cite{Marchini10} [SNPTESTv2], \cite{Purcell07} [PLINK v1.07]).
This workaround obviates the scaling of the RAM requirement with the number of SNPs in the sample, but it simultaneously renders multilocus methods (LD calculation, haplotype
and interaction testing, etc.) on imputed data impractical without intolerably many file access operations or currently unimplemented buffering schemes.

\section{Implementation}
In this paper, we present a method of compressing imputed GWAS data in RAM, analogous to how the data are commonly stored in compressed format on disk.  We make a standard assumption: probabilistic genotypes are reported from imputation software to some fixed, maximum precision.  This precision does not require the entire width of double- or float-precision data types to store.  For example, MACH (\cite{Li10}) and IMPUTE2 (\cite{Marchini10}) report results $\in \{0.000,0.001,\ldots,0.999,1.000\}$; BEAGLE (\cite{Browning09} reports probabilities with up to four decimals of precision.  We note in passing that even this level of precision is likely unnecessary for consistent results from regression, and thus while standard file formats have default precision settings, these can be dynamically changed for an improved compression ratio.

We choose to implement in-RAM data compression using a modified form of the DEFLATE (gzip) algorithm, which is not known to be license-encumbered.  Conceptually, similar libraries could be written using other compression algorithms.  A full description of DEFLATE is beyond the scope of this paper (see \cite{rfc1951}); here, we present a brief overview and pertinent modifications.

Briefly, DEFLATE compresses an input string by representing substrings that repeat along the data as (distance, length) pairs, which act
as backwards references to previous text strings.  It then represents the stream of such backwards references
and character literals (unsigned ASCII characters $\in[0,255]$) using two Huffman codes generated specifically for a block of data.
We use the same compression scheme, with the following modifications:


\begin{itemize}
\item Literals are defined not as characters of text input, but as integers $\in [0,10^{precision}+1]$; the additional code is for an end-of-block signal.
\item Input data are hashed as pairs of integers, not as triplets of characters.
\item Repeats can range in length from 2 to 257 literals (modified from $[3,258]$ in native gzip).
\end{itemize}

\begin{figure}[!tpb]
\centerline{\includegraphics[scale=0.55]{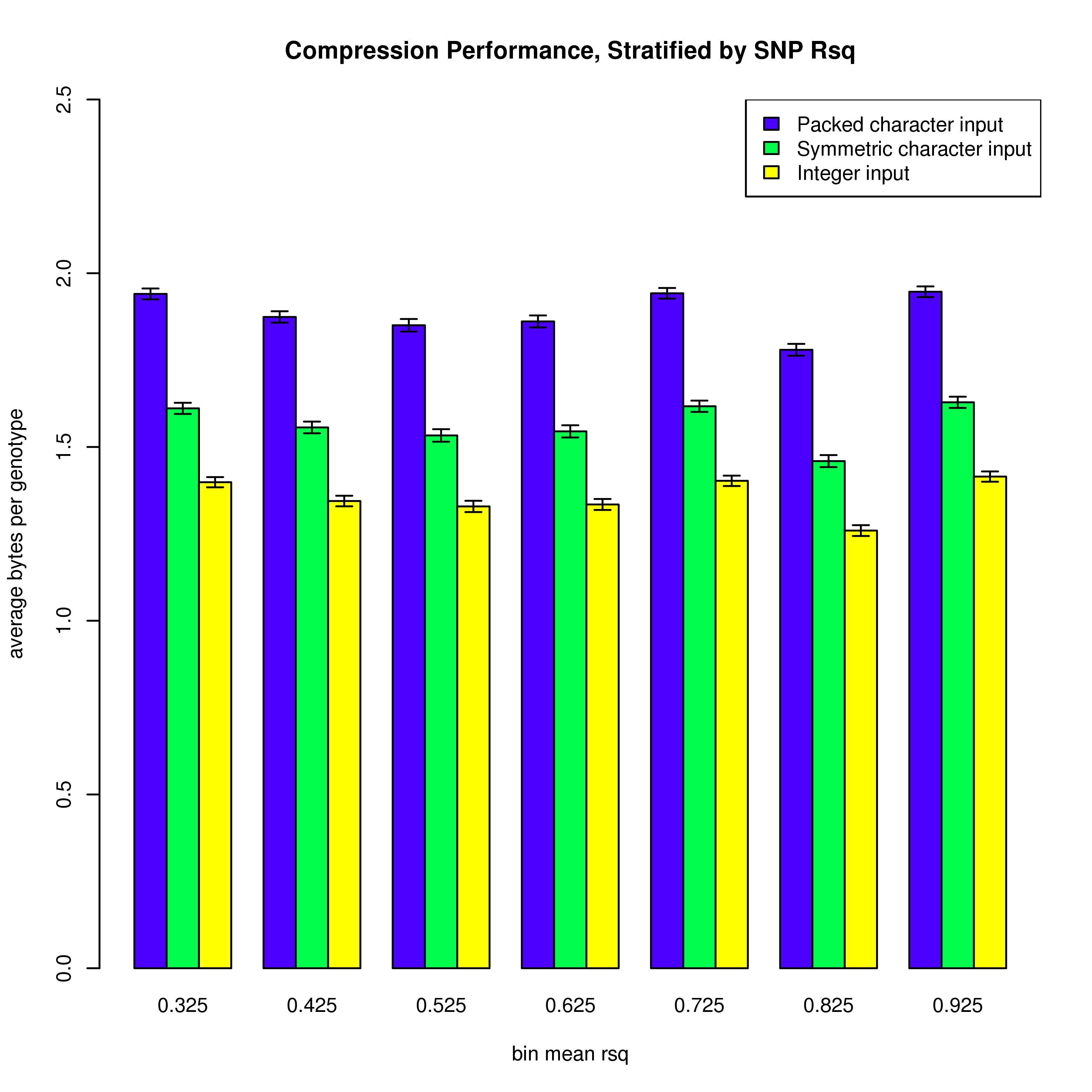}}
\caption{Average number of bytes RAM required per genotype (pair of probabilities), stratified by MACH rsq (interpreted as expected quality of imputation at the SNP; 0.3
minimum threshold for acceptable imputation; 1.0 max).  ``Packed character input:'' default gzip compression on characters; input probabilities packed
into 10 bits (20 bits per genotype) and stored with no padding.  ``Symmetric character input:'' default gzip compression on characters; input probabilities
packed into three characters (24 bits) as 12 bit values with leading MSB ``00''.  ``Integer input:'' modified gzip compression defined in this paper;
gzip operates on integer representations of probabilities on [0,1000].  Storing without compression leads to 16 bytes per type.  The improvement of compression with the modification of gzip suggests that maximizing the matches detected in the dataset, and minimizing the literals encountered, improves overall compression.}\label{fig:01}
\end{figure}

As genotype data are read, they are converted from decimals in $[0,1]$ with precision $p$ to integers in $[0,10^p]$, and presented to the algorithm.
We observe that these modifications to default gzip functionality lead to substantial benefits over na\"{i}ve compression using gzip or bzip2 (Figure ~\ref{fig:01}).

The version 1.0 release of this library natively supports common imputation file formats: MACH, IMPUTEv2, and BEAGLE; other formats may be specified using a standard format specification.  Optimal performance from the library is attained by iteration across a given dataset from front to back, without random access to variants; this is due to the partitioning of the dataset into blocks, and the need to unspool blocks from the beginning to access contained variants.  Random access is permitted in time asymptotic to the block size used to generate the data.  In practice, most statistical analyses in genome-wide scans simply access all genotype data from beginning to end, thus we do not anticipate this issue affecting most users.

\section{Conclusion}

We presented a C/C++ library for efficient, lossless in-memory storage of probabilistic genotypes.  Looking forward to maturation of the library, we hope that its existence
will encourage developers and statisticians to extend the current regime of imputed data analysis methods to more properly handle missingness in uncertain genotypes.

\section*{Acknowledgement}
The compression utility gzip is written and maintained by Jean-loup Gailly and Mark Adler (\cite{gzip}); although the algorithm is derived from the description of this implementation of DEFLATE, no source code from that project was used for this library.

\end{document}